\documentclass[sigconf,screen,authorversion]{acmart}
\usepackage{listings}

\def\finex{{\unskip\nobreak\hfil
\penalty50\hskip1em\null\nobreak\hfil$\diamond$
\parfillskip=0pt\finalhyphendemerits=0\endgraf}}

\lstdefinelanguage{JavaScript}{
  morekeywords=[1]{break, continue, delete, else, for, function, if, in,
    new, return, this, typeof, var, void, while, with, type, constructor, public},
morekeywords=[2]{false, null, true, boolean, number, undefined,
    Array, Boolean, Date, Math, Number, String, Object},
morekeywords=[3]{eval, parseInt, parseFloat, escape, unescape},
  sensitive,
  morecomment=[s]{/*}{*/},
  morecomment=[l]//,
  morecomment=[s]{/**}{*/}, morestring=[b]',
  morestring=[b]"
}[keywords, comments, strings]

\lstdefinelanguage[ECMAScript2015]{JavaScript}[]{JavaScript}{
  morekeywords=[1]{await, async, case, catch, class, const, default, do,
    enum, export, extends, finally, from, implements, import, instanceof,
    let, static, super, switch, throw, try},
  morestring=[b]` }
\lstalias[]{ES6}[ECMAScript2015]{JavaScript}

\definecolor{mediumgray}{rgb}{0.3, 0.4, 0.4}
\definecolor{mediumblue}{rgb}{0.0, 0.0, 0.8}
\definecolor{forestgreen}{rgb}{0.13, 0.55, 0.13}
\definecolor{darkviolet}{rgb}{0.58, 0.0, 0.83}
\definecolor{royalblue}{rgb}{0.25, 0.41, 0.88}
\definecolor{crimson}{rgb}{0.86, 0.8, 0.24}

\lstdefinestyle{JSES6Base}{
  basicstyle=\ttfamily\footnotesize\lstextra,
  breakatwhitespace=false,
  breaklines=false,
  columns=fullflexible,
  commentstyle=\color{mediumgray}\upshape,
  emph={},
  emphstyle=\color{crimson},
  extendedchars=true,  fontadjust=true,
  identifierstyle=\color{black},
  keepspaces=true,
  keywordstyle=\color{mediumblue},
  keywordstyle={[2]\color{darkviolet}},
  keywordstyle={[3]\color{royalblue}},
  numbers=left,
  numbersep=3pt,
  xleftmargin=8pt,
  numberstyle=\tiny\color{black},
  showlines=true,
  showspaces=false,
  showstringspaces=false,
  showtabs=false,
  stringstyle=\color{forestgreen},
  tabsize=2,
  upquote=true  }
\let\lstextra=\relax

\lstdefinestyle{JavaScript}{
  language=JavaScript,
  style=JSES6Base
}
\lstdefinestyle{ES6}{
  language=ES6,
  style=JSES6Base
}

\usepackage{tikz}
\usetikzlibrary{arrows,automata,calc}
\usetikzlibrary{shapes.symbols,shadows,decorations}
\tikzset{
  gt/.style={
         ->,
         >=stealth',
         shorten >=.1pt,
         auto,
         node distance=5cm,
         scale = 1,
         every state/.style={inner sep = 2pt, minimum size = 0pt, font=\footnotesize},
         transform shape
  },
lt/.style={
         ->,>=stealth',shorten >=1pt,auto,node distance=5cm,scale = 0.97,transform shape,
  },
  initial/.style={
         state,initial by arrow, initial text={}
  }
}

\title{Behaviorally Typed State Machines in TypeScript for Heterogeneous Swarms}



\author{Roland Kuhn}
\orcid{0000-0003-1582-6238}
\affiliation{\institution{Actyx AG}\city{M\"unchen}\country{Germany}}
\email{roland@actyx.io}

\author{Alan Darmasaputra}
\orcid{0009-0006-6002-2414}
\affiliation{\institution{Actyx AG}\city{M\"unchen}\country{Germany}}
\email{alan@actyx.io}

\setcopyright{acmlicensed}
\acmPrice{15.00}
\acmDOI{10.1145/3597926.3604917}
\acmYear{2023}
\copyrightyear{2023}
\acmSubmissionID{issta23demo-p3-p}
\acmISBN{979-8-4007-0221-1/23/07}
\acmConference[ISSTA '23]{Proceedings of the 32nd ACM SIGSOFT International Symposium on Software Testing and Analysis}{July 17--21, 2023}{Seattle, WA, USA}
\acmBooktitle{Proceedings of the 32nd ACM SIGSOFT International Symposium on Software Testing and Analysis (ISSTA '23), July 17--21, 2023, Seattle, WA, USA}
\received{2023-05-18}
\received[accepted]{2023-06-08}

\begin{document}

\begin{abstract}
  A heterogeneous swarm system is a distributed system where participants come and go, communication topology may change at any time, data replication is asynchronous and partial, and local agents behave differently between nodes.
  These systems are hard to design and reason about, mainly because we desire a particular class of behaviors to emerge from the interplay of heterogeneous individual agents.
  Nevertheless, mission-critical operations like manufacturing process orchestration in factories use such systems due to their uncompromising availability and resilience of computing services.

  This paper presents a set of TypeScript libraries to model peer-to-peer workflows as state machines, execute them using the Actyx middleware, and check the shape of these machines for conformance to a \emph{swarm protocol}.
  The swarm protocol describes an idealized global view of the cooperation of machines of different roles.
  It directly corresponds to a diagram a product manager would sketch on a whiteboard; this allows for verifying that the coded state machines correctly implement the product specification.
  A well-formed swarm protocol also guarantees that conforming machines will achieve \emph{eventual consensus} on the overall state progression even in the absence of further coordination.
  This tool is for developers of business logic for heterogeneous swarm systems, helping them verify that their protocols and implementations are correct.
  
  \textbf{Tool repo:} \url{https://github.com/Actyx/machines}
\end{abstract}

\maketitle

\section{Introduction}

Picture an automated factory shop floor: autonomous vehicles are buzzing to move materials, parts, and finished goods between machines and smart warehouses.
Participants of various kinds come and go dynamically and collaborate locally without centralized planning and oversight.
All communication is local and topology changes over time; groups form and dissolves around collaborative workflows; interruptions of participants or communication links are frequent, brief, and local---the rest of the swarm continues to function; participants may fail-stop or be newly introduced.
This setting is a \emph{heterogeneous swarm}, similar to \emph{local-first cooperation}~\cite{lfc}.

For parties who employ such systems, like programmers or product designers, the goal is to ensure that desired outcomes are eventually achieved; e.g.\,the factory shall produce a given set of items.
To prescribe the precise order and message types for communication between all participants is too rigid of an approach and curtails the needed concurrency.
Instead, we describe the structure of ephemeral collaboration within a smaller group of nodes so that such workflows proceed as designed; more formally, to ensure \emph{eventual consensus}~\cite{ec} on the overall workflow progression.

The tool presented in this paper aims to assist the process of designing, programming, and verifying such ephemeral collaboration.
For that, we apply the theory of \emph{swarm protocols} \cite{kmt23arxiv} using the Actyx~\cite{actyx} middleware; we improve the usability of the tool presented in that paper by moving from an external analysis of the TypeScript~\cite{typescript} code to a type-driven internal DSL, supporting IDE content assist as well as the behavioral checking within the usual unit testing cycle.

\begin{example}[Our running example]\label{ex:proto}
  A machine has finished its assigned production step and requests the fleet of logistics robots to pick up the workpieces.
  We use an auction mechanism illustrated below to select a logistics robot without requiring full oversight.

  \begin{center}
    \includegraphics{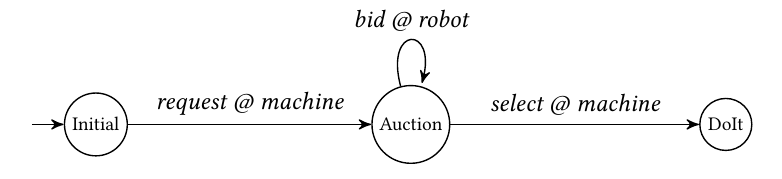}
  \end{center}

  \noindent
  First, the machine broadcasts a request (including the specifics like the location and the item being picked up).
  Then, available and capable robots place bids (accompanied by scoring data such as estimated time arrival and energy consumption).
  Finally, the machine selects one of the bids.
  \finex
\end{example}

The example demonstrates all essential features: sequential transitions, cycles, and choice without unique selector (notice that both the machine and all robots can concurrently act in the auction state).
In a real implementation, the workflow will be extended with timeouts, full logistic request execution tracking, or capturing on the protocol level which robot won the auction.
These extension are achievable with the tool's current feature set and will merely increase the size of the state machine.

\subsection{Brief overview of the theory}
\label{sec:theory}

Swarm protocols~\cite{kmt23arxiv} are used to formalize the communication model of the Actyx middleware: new \emph{events} are appended to a log owned by the local \emph{node}.
These logs are eventually replicated to other Actyx nodes.
A coordination-free total order of events is derived from a Lamport clock~\cite{lamport.clock} paired with a unique identifier for the emitting node; on this basis available logs are merged, sorted, and presented for local queries.
The paper then describes local agents as state machines that consume the locally available events to arrive at a state that may enable commands; events that are not expected in the current state are discarded.
Invoking a command is the process by which events are created and appended to the log, eventually leading to state updates throughout the swarm.

The second part of the paper introduces \emph{swarm protocols} as a description of how a set of machines works together to implement a workflow (like the one in Example~\ref{ex:proto}).
Such protocol is also described as a state machine---transitions labelled with command, role, and event log---assuming synchronous execution on all nodes to determine its intended semantics.
Each role can be played by any positive number of machines where the machine's state transition structure is prescribed to be equivalent to the projection of the protocol for the machine's role.
Since commands are enabled for specific roles and event types are filtered for each role by a \emph{subscription}, the resulting machines differ in shape across the swarm.

This asymmetry implies that machines of different roles may interpret different parts of the replicated and merged event log, coming to divergent conclusions on which branches the overall workflow has taken.
In other words, it is in general not the case that the logs produced by the non-coordinated execution of a set of correct machines will result in eventual consensus~\cite{ec} among the swarm.
This notion of consensus here means that there is a sequence of transitions of the swarm protocol that, when projected to the machine level, matches the local transition sequence as soon as sufficient log prefixes are replicated.
The main result of the paper is that for a swarm protocol satisfying a set of well-formedness conditions, every execution of correspondingly correct machines produces event logs that achieve eventual consensus.

\section{Overview of the Tool}

\lstMakeShortInline!

The tools presented in this paper are two TypeScript libraries available via \texttt{https://npmjs.org}:
\texttt{@actyx/machine-runner}~\cite{machineRunner} contains a DSL for describing a finite state machine as well as a facility for instantiating, evaluating, and interacting with such a machine, and
\texttt{@actyx/machine-check}~\cite{machineCheck} provides functions for verifying a swarm protocol's well-formed\-ness as well as validating a finite state machine against the swarm role it is supposed to play.

\subsection{Declaring machines}

Declaring a state machine proceeds by first creating event factories for every event type relevant to the swarm protocol.
These factories are used to construct events from a data payload and a type name, yielding a plain JavaScript object tagged with a !type! property.
The factory object also represents the event type at runtime, in particular for validating whether a given received event is of this type.

\begin{lstlisting}[name=main]
const requested = Event.design('requested')
  .withPayload<{ id: string; from: string; to: string }>()
const bid = Event.design('bid')
  .withPayload<{ robot: string; delay: number }>()
const selected = Event.design('selected')
  .withPayload<{ winner: string }>()
\end{lstlisting}
Next, event factories are bundled into a !SwarmProtocol!, which provides the context in which a state machine is declared for each participating role.

\begin{lstlisting}[name=main]
const transportOrderEvents = [requested, bid, selected] as const
const transportOrder = SwarmProtocol.make('transportOrder',
                                          transportOrderEvents)
const OrderForRobot = transportOrder.makeMachine('robot')
const TransportOrderForMachine = ...
\end{lstlisting}
Here, !OrderForRobot! is a builder for the state machine representing the ``robot'' role in our swarm protocol.
We proceed by first declaring its states, including their respective commands.

\begin{lstlisting}[name=main]
export const Initial = OrderForRobot.designState('Initial')(*@\label{state-start}@*)
  .withPayload<{ robot: string }>().finish()
export const Auction = OrderForRobot.designState('Auction')
  .withPayload<{ id: string; from: string; to: string;
                 robot: string; scores: Score[] }>()
  .command((*@\label{state-command-bid-declare}@*)'bid', [bid], 
    (ctx, delay: number) => [{ robot: ctx.self.robot, delay }](*@\label{state-command-bid-fn}@*))
  .finish()
export const DoIt = OrderForRobot.designState('DoIt')
  .withPayload<{ robot: string, winner: string }>.finish()(*@\label{state-end}@*)
\end{lstlisting}
Note how the command handler defined on line~\ref{state-command-bid-fn} constructs the emitted !bid! event payload from both the current state and the !delay! argument:
while the !ctx! argument is automatically passed in by the machine runner, the !delay! needs to be provided by the code invoking the command.
The final remaining piece now is the declaration of the state machine's transitions.
Each state can react to any number of event type sequences, keeping in mind that the first type in the list must be unique within that state since it selects the transition (i.e. we model deterministic state machines).

\begin{lstlisting}[name=main]
Initial.react([requested], Auction,(*@\label{machine-struct}@*)
  (ctx, r) => ({ ...ctx.self, ...r.payload, scores: [], }))(*@\label{machine-comp}@*)
Auction.react([bid], Auction, 
  (ctx, b) => {ctx.self.scores.push(b.payload); return ctx.self})
Auction.react([selected], DoIt, 
  (ctx, s) => ({robot: ctx.self.robot, winner: s.payload.winner}))
\end{lstlisting}
Line~\ref{machine-struct} and similar mark the structure of the machine, e.g.\ that a !requested! event may take it from the !Initial! into the !Auction! state, while line~\ref{machine-comp} and similar describe the computation of the target state payload, with !ctx! again being the current state passed in by the machine runner.

Take note that the state machine requires !requested! and !selected! events to progress from state !Initial! to state !DoIt!, but there are no corresponding commands emitting these event types.
As per the state diagram in example~\ref{ex:proto} these events shall be emitted by the ``machine'' role, so they will show up in commands declared on !TransportOrderForMachine! once that is implemented; this is the heterogeneity that our tool natively supports.

One noteworthy aspect of the API demonstrated above is that it clearly separates the persistent data model (i.e. the event types) from the ephemeral in-memory data model (i.e. the state payload).
This allows the user to place related runtime information in the ephemeral state, as shown with the !robot! property that identifies the current machine within its role;
an alternative model would remove this property and require the robot name to be passed into the !bid! command handler.

\subsection{Executing machines}

The code below shows how the application code instantiates a state machine.
A call is made to !createMachineRunner! with a connection to Actyx, event tags to uniquely identify this workflow instance (i.e.\,\emph{protocol session}), the initial state, and its corresponding payload.
The value !'4711'! on line ~\ref{tags} is the identifier of this workflow.
The value !'agv1'! assigned to the !robot! property is the identifier of the robot; take note that the last argument of !createMachineRunner! matches the type of the payload of !Initial! defined previously.
\begin{lstlisting}[name=main]
const manifest = {appId: 'acm', displayName: 'acm', version: '1'}
const conn = await Actyx.of(manifest) // connect to local Actyx
const tags = transportOrder.tagWithEntityId('4711')(*@\label{tags}@*)
const machine =
  createMachineRunner(conn, tags, Initial, { robot: 'agv1' })
\end{lstlisting}

The application program interacts with the swarm through the state machine by observing its state and invoking the offered commands.
The returned state machine instance !machine! in the code above is a JavaScript !AsyncIterable! which allows interaction via the !for await (...)! loop as shown in line ~\ref{for-await-loop} below.
State values such as !const state! in line ~\ref{for-await-loop} are opaque objects describing the union of all the machine's possible states.
\begin{lstlisting}[name=main]
  let IamWinner = false
  for await (const state of machine) {(*@\label{for-await-loop}@*)
    if (state.is(Auction)) {(*@\label{downcast-auction}@*)
      const open = state.cast()
      const payload = open.payload
      if (!payload.scores.find((s) => s.robot === payload.robot)){(*@\label{guard}@*)
        await open.commands?.bid(1)(*@\label{command}@*)
      }
    } else if (state.is(DoIt)) {(*@\label{downcast-doit}@*)
      const assigned = state.cast().payload
      IamWinner = assigned.winner === assigned.robot(*@\label{determine}@*)
      if (!IamWinner) break
      // now we have the order and can start the mission
    }
  }
\end{lstlisting}
To observe or interact, the object needs to be downcast into a desired state using !.is(...)! and !.cast(...)! as shown in lines ~\ref{downcast-auction} and ~\ref{downcast-doit}, providing type-safe access to payload and commands.
F.e.\,the command invocation on line ~\ref{command} results in a !bid! event, whose payload is calculated by the function shown on line ~\ref{state-command-bid-declare} and the argument provided by the caller; the event is then sent to Actyx for publication.
Event publishing is asynchronous, hence we !await! the successful result of invoking the !bid()! command---if the local Actyx node weren't running then this would result in a runtime exception.
Published events are then returned via the standing event subscription for the machine and consumed via the reaction handlers, possibly leading to state transitions.
We decided to design the system so that unexpected events (i.e.\ not matching any transition) are discarded and passed to an optional hook; this hook allows the application to compensate for previous actions that have been invalidated by eventual consensus.
Commands are dynamically disabled after invocation as well as while waiting for further events to complete a multi-event state transition, hence the !?.! call syntax.

The async loop iterates when there is an incoming state transition, but only after the previous iteration, if any, has finished executing.
The !state.is(Auction)! and !state.is(DoIt)! branches depict how the robot (i.e.\,the application) decides what to do when it observes the state after said transitions.
The aforementioned command call is the robot placing a bid for the transport request, essentially interacting with the swarm.

\subsection{Analyzing swarm protocols and machines}

The machines declared above contain all needed information to apply our theory \cite{kmt23arxiv} (cf.\,Section~\ref{sec:theory}): the machine builder DSL stores the states, commands, and event transitions corresponding to the syntactic elements of machine types.
This allows the machine check library to provide two essential verification steps, ideally used within the application code test suite.
The first checks if the swarm protocol given the event subscription implemented in the code achieves eventual consensus without further coordination.
The second checks whether the TypeScript state machine code correctly implements its assigned role in the swarm protocol.

\begin{lstlisting}[name=main]
const protocol = {(*@\label{decl-start}@*) initial: 'Initial', transitions: [
  { source: 'initial', target: 'auction'
    label: {cmd: 'request', logType:['requested'], role:'machine'},
  }, {
    source: 'auction', target: 'auction'
    label: {cmd: 'bid', logType:['bid'], role: 'robot'},
  }, {
    source: 'auction', target: 'doIt'
    label: {cmd: 'select', logType:['selected'], role: 'machine'},
  } ]}(*@\label{decl-end}@*)
const robotShape =
  OrderForRobot.createJSONForAnalysis(Initial)(*@\label{machine-json}@*)
const machineShape = ...
const subs = {
  robot: robotShape.subscriptions,
  machine: machineShape.subscriptions
}
expect(checkSwarmProtocol(protocol, subs))
  .toEqual({ type: 'OK' })
expect(checkProjection(protocol, subs, 'robot', robotShape))
  .toEqual({ type: 'OK' })
\end{lstlisting}

On lines~\ref{decl-start}--\ref{decl-end} above we declare the workflow structure corresponding to Example~\ref{ex:proto}; one array item corresponds to one transition arrow with source, target, and a label detailing role, command name, and emitted event log type.
Line~\ref{machine-json} demonstrates how to extract the machine type from the machine implementation, yielding both the state machine graph and the inferred event subscriptions.
We then use !checkSwarmProtocol! to verify the well-formedness of the protocol under the implemented subscription, which guarantees eventual consensus.
If that succeeds, !checkProjection! computes the machine projection from the swarm protocol and compares it to the extracted state machine graph from TypeScript code; this structural comparison verifies that the implemented transition graph will process sequences of event types in the same way as the projected one, and that it enables the same set of commands in each visited state.
Any discrepancy will be listed in an !'ERROR'! result variant, giving the programmer a clear indication of what is wrong.

\section{Validation of the Approach}

The tools presented above are currently being used and evaluated in the context of the EU HorizonEurope project ``TaRDIS''.
As part of the TaRDIS deliverables we will assess how developer productivity, confidence, and defect rate are improved over previously used tools, including centralized approaches using for example MQTT \cite{mqtt} and relational databases, or the Actyx Pond library~\cite{pond} which uses the same underlying event dissemination technology as this tool.

This assessment will be carried out using a case study and a set of experiments.
The case study has already started at a machine and plant building company and involves several of that company's developers.
After having implemented a working proof of concept using the Actyx Pond library (which requires manual checking of protocol correctness) they reimplement their peer-to-peer workflows using this tool, allowing comparative measurements of implementation time, effort, and defect rate.
Further, they will implement new features (i.e.\ protocol changes) and compare the time and effort of deploying those in production with previous experience on modifying similar systems.
In addition to the case study we plan to conduct experiments with groups of students at the institutes participating in TaRDIS, asking them to implement swarm participants conforming to given protocols using this tool and popular alternatives, as well as extending swarm protocols according to a feature specification.
This will allow us to measure time, effort, and defect rate as well as surveying non-functional aspects like developer confidence, satisfaction, and subjective efficacy.

The implementation work has been ongoing for several months already, we can preliminarily say that the more structured approach (compared to hand-rolled partial state updates in response to events) has led to productivity gains.
The subjective confidence of the developers in the correctness of the written source code has increased dramatically.
The latter is not unexpected since the application code previously needed to manually decide when to apply an event that arrived out-of-order and when to ignore or compensate it.

Another result of the ongoing validation is that while our approach enables the programmer to recognize conflicts and if necessary compensate them, a more convenient and safer API would be desirable for handling all invalidated events of a discarded branch at once; currently each invalidated event is handled individually.

\section{Related work}\balance

Our work improves upon the established Actyx Pond library, whose ``fish'' abstraction only permits symmetric business logic on all participants and requires the user to handle all events at all times.

Compared to the currently existing TypeScript libraries for formulating state machines (e.g. XState~\cite{xstate}), in addition to having the essential features (i.e state and transitions declaration and change observation), our tool focuses on enabling the construction of distributed state machines with built-in mechanisms for eventual-consistency.
In addition to supporting behavioral type checking, our machine declaration DSL provides a very high level of type safety, e.g.\ capturing declared commands per state and thus avoiding invalid actions; high quality type definitions also enable high fidelity content assistance by the IDE, showing correct and detailed completions and error messages.

In this brief paper we omit a comparison to \emph{state machine replication} (like Paxos or Raft) because it is well-known~\cite{CAP,FLP} that strongly consistent systems become unavailable under network partitions whereas we aim for perfect availability.

\section{Conclusion}

We present two TypeScript libraries to aid developers in implementing heterogeneous swarm systems, in particular regarding adhering to designed protocols between participants and in ensuring that the designed protocols will achieve eventual consensus in the absence of further coordination.
The tools are currently used to implement the orchestration of manufacturing processes on the factory shop floor but are not specific to this application domain---they are instead tailored to heterogeneous swarm systems in general.

Ascertaining the emergence of the desired overall behavior from an uncoordinated swarm system is an unsolved problem.
We carve out the smaller problem of analyzing the asymmetric cooperation on non-adversarial workflows to validate correctness properties, projecting them down onto local agent code and helping the developer correctly implement and confidently modify such code.

Emphasis has been placed on making full use of the capabilities of TypeScript to provide good error messages as well as precise context-sensitive completions in an IDE.
The representation of swarm protocols and state machines is geared towards integration with graphical editors---for coders and managers---and users are free to implement their own behavioral type checking schemes.

\begin{acks}
  This work is partially funded by the European Union (TaRDIS, 101093006).
  We thank Jos\'e Duarte and the anonymous reviewers.
\end{acks}

\bibliographystyle{ACM-Reference-Format}
\bibliography{bib}


\begin{thebibliography}{13}


\ifx \showCODEN    \undefined \def \showCODEN     #1{\unskip}     \fi
\ifx \showDOI      \undefined \def \showDOI       #1{#1}\fi
\ifx \showISBNx    \undefined \def \showISBNx     #1{\unskip}     \fi
\ifx \showISBNxiii \undefined \def \showISBNxiii  #1{\unskip}     \fi
\ifx \showISSN     \undefined \def \showISSN      #1{\unskip}     \fi
\ifx \showLCCN     \undefined \def \showLCCN      #1{\unskip}     \fi
\ifx \shownote     \undefined \def \shownote      #1{#1}          \fi
\ifx \showarticletitle \undefined \def \showarticletitle #1{#1}   \fi
\ifx \showURL      \undefined \def \showURL       {\relax}        \fi
\providecommand\bibfield[2]{#2}
\providecommand\bibinfo[2]{#2}
\providecommand\natexlab[1]{#1}
\providecommand\showeprint[2][]{arXiv:#2}

\bibitem[Actyx{ }AG(2022)]%
        {actyx}
\bibfield{author}{\bibinfo{person}{Actyx{ }AG}.}
  \bibinfo{year}{2020-2022}\natexlab{}.
\newblock \bibinfo{title}{Actyx developer website}.
\newblock
\newblock
\urldef\tempurl%
\url{https://developer.actyx.com}
\showURL{%
\tempurl}
\newblock
\shownote{accessed 2023-05-18}.


\bibitem[Actyx{ }AG(2023a)]%
        {pond}
\bibfield{author}{\bibinfo{person}{Actyx{ }AG}.}
  \bibinfo{year}{2020–2023}\natexlab{a}.
\newblock \bibinfo{title}{Actyx Pond library}.
\newblock
\newblock
\urldef\tempurl%
\url{https://www.npmjs.com/package/@actyx/pond/v/3.4.0}
\showURL{%
\tempurl}


\bibitem[Actyx{ }AG(2023b)]%
        {machineCheck}
\bibfield{author}{\bibinfo{person}{Actyx{ }AG}.}
  \bibinfo{year}{2023}\natexlab{b}.
\newblock \bibinfo{title}{@actyx/machine-check library}.
\newblock
\newblock
\urldef\tempurl%
\url{https://www.npmjs.com/package/@actyx/machine-check/v/0.2.0}
\showURL{%
\tempurl}


\bibitem[Actyx{ }AG(2023c)]%
        {machineRunner}
\bibfield{author}{\bibinfo{person}{Actyx{ }AG}.}
  \bibinfo{year}{2023}\natexlab{c}.
\newblock \bibinfo{title}{@actyx/machine-runner library}.
\newblock
\newblock
\urldef\tempurl%
\url{https://www.npmjs.com/package/@actyx/machine-runner/v/0.3.1}
\showURL{%
\tempurl}


\bibitem[Fischer et~al\mbox{.}(1985)]%
        {FLP}
\bibfield{author}{\bibinfo{person}{Michael~J. Fischer},
  \bibinfo{person}{Nancy~A. Lynch}, {and} \bibinfo{person}{Michael~S.
  Paterson}.} \bibinfo{year}{1985}\natexlab{}.
\newblock \showarticletitle{Impossibility of Distributed Consensus with One
  Faulty Process}.
\newblock \bibinfo{journal}{\emph{J. ACM}} \bibinfo{volume}{32},
  \bibinfo{number}{2} (\bibinfo{date}{apr} \bibinfo{year}{1985}),
  \bibinfo{pages}{374–382}.
\newblock
\showISSN{0004-5411}
\urldef\tempurl%
\url{https://doi.org/10.1145/3149.214121}
\showDOI{\tempurl}


\bibitem[Gilbert and Lynch(2002)]%
        {CAP}
\bibfield{author}{\bibinfo{person}{Seth Gilbert} {and} \bibinfo{person}{Nancy
  Lynch}.} \bibinfo{year}{2002}\natexlab{}.
\newblock \showarticletitle{Brewer's conjecture and the feasibility of
  consistent, available, partition-tolerant web services}.
\newblock \bibinfo{journal}{\emph{SIGACT News}} \bibinfo{volume}{33},
  \bibinfo{number}{2} (\bibinfo{date}{June} \bibinfo{year}{2002}),
  \bibinfo{pages}{51--59}.
\newblock
\showISSN{0163-5700}
\urldef\tempurl%
\url{https://doi.org/10.1145/564585.564601}
\showDOI{\tempurl}


\bibitem[Kuhn(2021)]%
        {lfc}
\bibfield{author}{\bibinfo{person}{Roland Kuhn}.}
  \bibinfo{year}{2021}\natexlab{}.
\newblock \bibinfo{title}{Local-First Cooperation}.
\newblock
\newblock
\urldef\tempurl%
\url{https://www.infoq.com/articles/local-first-cooperation/}
\showURL{%
\tempurl}


\bibitem[Kuhn et~al\mbox{.}(2023)]%
        {kmt23arxiv}
\bibfield{author}{\bibinfo{person}{Roland Kuhn}, \bibinfo{person}{Hern\'an
  Melgratti}, {and} \bibinfo{person}{Emilio Tuosto}.}
  \bibinfo{year}{2023}\natexlab{}.
\newblock \bibinfo{title}{Behavioural Types for Local-First Software}.
\newblock
\newblock
\showeprint[arxiv]{2305.04848}~[cs.DC]
\newblock
\shownote{To appear in ECOOP}.


\bibitem[Lamport(2019)]%
        {lamport.clock}
\bibfield{author}{\bibinfo{person}{Leslie Lamport}.}
  \bibinfo{year}{2019}\natexlab{}.
\newblock \bibinfo{booktitle}{\emph{Time, Clocks, and the Ordering of Events in
  a Distributed System}}.
\newblock \bibinfo{publisher}{Association for Computing Machinery},
  \bibinfo{address}{New York, NY, USA}, \bibinfo{pages}{179–196}.
\newblock
\showISBNx{9781450372701}
\urldef\tempurl%
\url{https://doi.org/10.1145/3335772.3335934}
\showDOI{\tempurl}


\bibitem[Light(2017)]%
        {mqtt}
\bibfield{author}{\bibinfo{person}{R.~A. Light}.}
  \bibinfo{year}{2017}\natexlab{}.
\newblock \showarticletitle{Mosquitto: server and client implementation of the
  MQTT protocol}.
\newblock \bibinfo{journal}{\emph{The Journal of Open Source Software}}
  \bibinfo{volume}{2}, \bibinfo{number}{13} (\bibinfo{year}{2017}).
\newblock
\urldef\tempurl%
\url{https://doi.org/10.21105/joss.00265}
\showDOI{\tempurl}


\bibitem[Microsoft(2023)]%
        {typescript}
\bibfield{author}{\bibinfo{person}{Microsoft}.}
  \bibinfo{year}{2012-2023}\natexlab{}.
\newblock \bibinfo{title}{TypeScript: JavaScript with Syntax for Types}.
\newblock
\newblock
\urldef\tempurl%
\url{https://www.typescriptlang.org/}
\showURL{%
\tempurl}


\bibitem[STATELY(2023)]%
        {xstate}
\bibfield{author}{\bibinfo{person}{STATELY}.} \bibinfo{year}{2023}\natexlab{}.
\newblock \bibinfo{title}{XState website}.
\newblock
\newblock
\urldef\tempurl%
\url{https://xstate.js.org/docs/}
\showURL{%
\tempurl}
\newblock
\shownote{accessed 2023-05-18}.


\bibitem[Tseng(2019)]%
        {ec}
\bibfield{author}{\bibinfo{person}{Lewis Tseng}.}
  \bibinfo{year}{2019}\natexlab{}.
\newblock \showarticletitle{Eventual Consensus: Applications to Storage and
  Blockchain}. In \bibinfo{booktitle}{\emph{2019 57th Annual Allerton
  Conference on Communication, Control, and Computing (Allerton)}}.
  \bibinfo{pages}{840--846}.
\newblock
\urldef\tempurl%
\url{https://doi.org/10.1109/ALLERTON.2019.8919675}
\showDOI{\tempurl}


\end{thebibliography}

\end{document}